\documentclass{optica-article}

\journal{opticajournal} 

\articletype{Research Article}

\usepackage{lineno}

\usepackage{physics}
\usepackage{cancel}
\usepackage[english]{babel}

\newcommand{\FI}{\mathcal{J}} 
\newcommand{\QFI}{\mathcal{I}}
\newcommand{\Np}{p} 
\newcommand{\poi}{\vartheta} 
\newcommand{\nuis}{\xi} 
\newcommand{\schur}[2]{{#1 #1\cdot #2}}

\DeclareMathOperator{\Var}{Var}
\DeclareMathOperator{\Cov}{Cov}

\newcommand{\appendixnumbering}{
    \setcounter{section}{0}
    \renewcommand{\thesection}{S\arabic{section}}
    \renewcommand{\thesubsection}{\thesection.\arabic{subsection}}
}

\newcommand{\titlefont}{\normalfont\sffamily\bfseries\fontsize{22}{25}\selectfont}
\newcommand{\startappendix}[1]{
    {\raggedright \titlefont #1\par}%
    \vskip11pt
    \appendixnumbering
}

\begin{document}

\title{Multiparameter Maximum Information States for Coherent Diffraction Measurements}

\author{Bram Verreussel,\authormark{1,*} Jacob Seifert,\authormark{1} and Allard P. Mosk\authormark{1}}

\address{
\authormark{1}Nanophotonics, Debye Institute for Nanomaterials Science and Center for
Extreme Matter and Emergent Phenomena, Utrecht University, P.O. Box 80000,
Utrecht 3508 TA, The Netherlands    
}

\email{\authormark{*}b.verreussel@uu.nl} 


\begin{abstract*}
    In metrology, Fisher information is an important metric that quantifies the precision that can be achieved in a measurement. For optical measurements using coherent light it has been shown that Fisher information can be expressed simply using the scattering matrix of the system. Fisher information can be maximized over the input modes to achieve \textit{maximum information states}, which produce optimally precise estimates for a parameter when the system is limited by photon noise. Here, we extend this approach to multiparameter estimation, in which case Fisher information takes the form of a matrix. We consider several scalar functions of the Fisher matrix to optimize the precision in multiple parameters at the same time. We also consider strategies for dealing with nuisance parameters, which can degrade the achievable precision of other parameters but are not of interest to measure. We corroborate our findings numerically using a scattering system of 2D coupled dipoles.

\end{abstract*}

\section{Introduction}
Optical metrology entails the various methods of using light to perform measurements. Light is ideal for metrology since it is fast, non-invasive and possesses many degrees of freedom to exploit. In particular, the spatial degrees of freedom can be controlled using a Spatial Light Modulator (SLM) to produce arbitrary wavefronts. In wavefront shaping, this is used to focus intensity even inside a complex scattering environment \cite{vellekoop_focusing_2007, pai_scattering_2021, horstmeyer_guidestar-assisted_2015, yoon_deep_2020, ammar_upper_2025, rotter_light_2017}. This can be used to direct the intensity to the regions of interest to increase precision.
More recently, attention has shifted to directly increasing the information content of the light. In particular, Bouchet et al. \cite{bouchet_maximum_2021} have shown that it is possible to construct \textit{maximum information states}, which are light fields with provably maximum information about some parameter of interest.

When light interacts with an object, the information about that object is imprinted and carried away to the detector\cite{hupfl_continuity_2024}. By shaping the field, the amount of information that is extracted from the object and carried to the detector can be greatly increased. In particular, parts of the field with high intensity do not necessarily have high information \cite{weimar_controlling_2025}. By increasing the information efficiency it is possible to reduce measurement resources such as exposure time and optical power, which is useful when the optical power is limited by a damage threshold or when the detector is close to saturation. 
Also of interest is the fact that information is locally conserved in absence of absorption, even in complex scattering environments\cite{hupfl_continuity_2024}. This makes information an attractive target for wavefront shaping. 

The relevant metric for precision measurements is \textit{Fisher information} (FI). Estimation theory posits that any experimental determination of a parameter always happens through measurement of some stochastic, or noisy, variable\cite{cover_elements_2005, kay_fundamentals_2013}. Noise in the measurements then translates to noise in the parameter estimate. Fisher information puts a hard lower bound on the variance of the parameter estimate, which is known as the Cram\'er-Rao Lower Bound (CRLB). If Fisher information is increased, the lower bound will be relaxed and this will allow for estimations with lower variance. This motivates maximizing Fisher information for maximal precision. 

Much of previous work has been focused on single parameter estimation. However, practical applications often require the estimation of multiple parameters. In multiparameter FI, the FI takes the form of a matrix whose inverse provides a lower bound on the covariance matrix. Thereby not only capturing the information of each individual parameter, which are located on the diagonal elements of the matrix, but also the ``cross information'' on the off-diagonals between different parameters which can negatively impact estimations.  The Fisher matrix has shown widespread use and it appears in many domains:  cosmology\cite{heavens_generalised_2014, elsner_likelihood_2012}, oceanography\cite{nolet_statistical_1995} , sensor placement\cite{kammer_sensor_1991} and neural networks\cite{pascanu_revisiting_2013, weimar_fisher_2025}  to name just a few. 

Previous work has shown results for FI in terms of the \textit{intensity} of the output field\cite{ de_graaff_very_2025, xi_information-efficient_2020, feldman_nanometer_2024, feldman_information_2025}. In particular, Bouchet et al. \cite{bouchet_influence_2020, bouchet_optimizing_2021} studied this in the context of multiparameter Fisher information, but multiparameter FI using the scattering matrix has yet to be investigated. In this work, we introduce such an expression. This approach conveys two important benefits. Firstly, measuring the scattering matrix of the system assumes no prior knowledge of the system, which makes it possible to do model free optimization. Secondly, by incorporating phase in the measurement, the information that can be extracted from the system increases\cite{bouchet_maximum_2021}. 

Once we have the Fisher matrix, we have to choose a metric to optimize for. Making this choice is the essence of the field of \textit{optimal experiment design}\cite{huan_optimal_2024}. In this work, we will discuss a number of these metrics, better known as \textit{optimality criteria}, and evaluate their performance on a numerical model of coupled dipoles. We will also investigate a number of ways to deal with \textit{nuisance parameters}, parameters that are necessary to account for, but whose exact value is uninteresting to the experimenter. Altogether, this creates a useful framework to do precise multiparameter estimation using the scattering matrix of the system.

 \section{Theory}
    \subsection{Maximum Information States}
For coherent states it has been shown that the Fisher information with respect to some parameter $\theta$ is given by\cite{bouchet_maximum_2021}
\begin{align}
   \FI(\theta)=4\sum_k\abs{\partial_{\theta} E_k^{\mathrm{out}}}^2,
\end{align}
where $E_k^{\mathrm{out}}$ represents the amplitude of the $k$-th mode in the output field and $\partial_\theta$ is the partial derivative with respect to $\theta$. Loosely speaking, higher Fisher information will allow for more precise estimations of the parameter. To determine which input states maximize this expression, we can relate the output field to the input field using the scattering matrix, $E^{\mathrm{out}}=SE^{\mathrm{in}}$, which is a complete description of the system when in the linear optics regime. In that case the Fisher information is written as
\begin{equation}
    \FI(\theta)=4\bra{E^{in}}\partial_{\theta}S^\dagger \partial_{\theta}S\ket{ E^{\mathrm{in}}}.
\end{equation}
The operator $F\equiv \partial_{\theta}S^\dagger \partial_{\theta}S$ is called the Fisher information operator. It is Hermitian and consequently, this expression can be maximized by performing an eigendecomposition on $F$ and finding the state with the largest eigenvalue, which are referred to as \textit{maximum information states}. When the sample is illuminated with such a state, the output state contains maximum information about the parameter of interest and this allows one to find estimators that are as precise as possible, within a specific photon budget.

    \subsection{Multiparameter Maximum Information States}
Here, we extend this framework to multiple parameters by expressing the Fisher information as
\begin{equation}
    \FI_{\mu\nu}=2\bra{E^{in}}(\partial_{\mu}S^\dagger \partial_{\nu}S +\partial_{\nu}S^\dagger \partial_{\mu}S )\ket{ E^{\mathrm{in}}},\label{eq:MPFI}
\end{equation}
where the Greek indices $\mu,\nu$ run over the different parameters $ \theta=(\theta_1,\dots,\theta_\Np)^T$, and $\partial_\mu$ is shorthand for $\pdv[]{\theta_\mu}$ . A derivation can be found in appendix \ref{ap:MQFI} and \ref{ap:MFI}. Here, $\FI_{\mu\nu}$ forms a $p$ by $p$ matrix. In brackets we recognize the multiparameter Fisher tensor $F_{\mu\nu}\equiv \partial_{\mu}S^\dagger \partial_{\nu}S +\partial_{\nu}S^\dagger \partial_{\mu}S $ which has size $p\times p\times N^{\mathrm{in}}\times N^{\mathrm{in}}$, where $N^{\mathrm{in}}$ is the number of input modes that are considered. From the form of \eqref{eq:MPFI}, it is clear that the Fisher tensor is symmetric in its parameter indices, i.e. $F_{\mu\nu}=F_{\nu\mu}$, and Hermitian in its mode indices, i.e. $F_{\mu\nu}^\dagger=F_{\mu\nu}$. The resulting Fisher matrix is positive semi-definite for any input state $\ket{E^{\mathrm{in}}}$, as one should expect. 

We should note that the prefactor in equation \eqref{eq:MPFI}, although theoretically possible, is in practice often replaced by a factor 1. This holds for example for digital off-axis holography. For more discussion on this prefactor we refer to appendix \ref{ap:MFI}. Throughout this document we use the prefactor 2.
    
\subsection{Fisher Information}
We assume the outcome of a measurement is described by a stochastic variable $X$ with probability distribution $p(X;\theta)$, where $\theta$ is the parameter that is to be estimated. The importance of Fisher information is exemplified by the Cram\'er-Rao bound, which for one parameter reads
\begin{equation}
    \Var(\hat\theta)\geq \frac{1}{\FI(\theta)} \label{eq:scalar-CR},
\end{equation}
where $\hat\theta(X)$ is any unbiased estimator and $\FI(\theta)$ is the Fisher information\footnote{In the context of precision measurements, we assume the a priori spread in the parameter value is negligible compared to the spread in $X$. }, given by\cite{cover_elements_2005}
\begin{equation}
    \FI(\theta)=E\bqty{\pqty{\pdv{}{\theta}\ln p(X;\theta) }^2}.
\end{equation} An estimator is \textit{any} function that estimates the parameter from measurements. An estimator is unbiased if its expectation value is equal to the true value of the parameter. The Cramér-Rao bound tells us that increasing Fisher information on a parameter allows to find estimators with higher precision, although their existence is not guaranteed. An estimator that achieves equality is called \textit{efficient}.

For multiple parameters, the Fisher information is encoded in the Fisher information matrix, given by
\begin{equation}
    \FI_{\mu\nu}=\mathbb E\bqty{\pqty{\partial_\mu \ln p(X;\theta)}  \pqty{\partial_\nu \ln p(X;\theta)} }. \label{eq:fisher-information-matrix}
\end{equation}
The Cram\'er-Rao Lower Bound is generalized to
\begin{equation}
    \Sigma\succeq \FI^{-1}(\theta) \label{eq:matrix-CR},
\end{equation}
where \begin{enumerate}
    \item $\FI^{-1}(\theta)$ is the inverse of the Fisher information matrix,
    \item $\Sigma_{\mu\nu}=\Cov(\hat\theta_\mu,\hat \theta_\nu)$ is the covariance matrix of estimators,
    \item the symbol $\succeq$ is used to indicate that the matrix $\Sigma-\FI^{-1}(\theta)$ is positive semi-definite, i.e. it has only positive or zero eigenvalues.
\end{enumerate}

Of particular importance are the diagonals of the inverse Fisher information matrix, which satisfy the following inequality\cite{kay_fundamentals_2013}:
\begin{equation}
    \Var(\hat\theta_\mu)\geq \FI_{\mu\mu}^{-1}.
\end{equation}
In other words, the diagonals of the inverse Fisher matrix form the Cram\'er-Rao bounds in the multiparameter setting. We also have the inequality
\begin{equation}
    \FI_{\mu\mu}^{-1}\geq \frac{1}{\FI_{\mu\mu}}.\label{eq:jcrlb-vs-scrlb}
\end{equation}
On the righthand side of equation \eqref{eq:jcrlb-vs-scrlb} we recognize the single parameter Cram\'er-Rao bounds. To distinguish these two bounds, we will refer to $1/\FI_{\mu\mu}$ as the \textit{scalar Cram\'er-Rao bound} and to $\FI^{-1}_{\mu\mu}$ as the \textit{joint Cram\'er-Rao bound}. The difference between the scalar and joint lower bound is caused by off-diagonal terms in the Fisher matrix. When the off-diagonals are zero, equation \eqref{eq:jcrlb-vs-scrlb} becomes an equality. Therefore, an ideal Fisher matrix has large values on the diagonals, and small or zero values on the off-diagonals. Note that when multiple parameters are present, only the joint Cram\'er-Rao bounds are meaningful. The scalar bound is naive, or overly optimistic. However, it represents the joint bound that \textit{could} be achieved if the system could somehow rid itself of all its cross-information.
    \subsection{Optimality Criteria\label{sec:optimality-criteria}}
Optimizing the Fisher information matrix for some design goal is central to the field of optimal experimental design. A practical solution is to encode your design goal into a scalar function of the Fisher information matrix, which can then be maximized or minimized. Three optimality criteria that are common from literature are D-, A- and E-optimality\cite{huan_optimal_2024}. We will discuss and then test how these optimality criteria perform in the context of multiparameter maximum information states. We also introduce another optimality criterion, the normalized trace, which has as main benefit that it can be optimized as an eigenvalue problem.

\subsubsection*{D-, A- and E-optimality\label{sec:DAE-optimality}}
The D-, A- and E-optimalities are defined as scalar functions of the Fisher information matrix. In D-optimality, the determinant of the Fisher matrix is maximized. In A-optimality, the trace of the inverse matrix is minimized, or the "average" of the joint CRLBs. In E-optimality, the smallest eigenvalue of the Fisher matrix is maximized; or equivalently, the largest eigenvalue of the inverse Fisher matrix is minimized. These three criteria can be seen as instances of the power mean acting on the eigenvalues of the Fisher matrix\footnote{Two optimality criteria are considered equal when they produce the same optimal states, i.e. when you can write one criteria as a monotonic function of the other.}, which is given by
\begin{equation}
M_p(\lambda_1,\dots,\lambda_n)=\left(\frac 1 n\sum_i \norm{\lambda_i}^p\right)^{1/p}   . 
\end{equation}
D-, A- and E-optimality then correspond to maximizing this expression for $p=0, -1, -\infty$ respectively. An optimization with $p$ large and positive is a "winner takes all" approach. It favors states where one eigenvalue is as large as possible at the cost of every other parameter. In contrast, an optimization with $p$ large and negative is an egalitarian approach. It favors states where all eigenvalues are as equal as possible. One could argue that an approach that is too much "winner takes all" is not so useful, since you are essentially performing single parameter optimization in an arbitrary direction in parameter space. This last fact might explain why optimality criteria with $p\geq 1$ are not common in literature.

It is also interesting to point out that D-optimality is the only criterion in this family that is invariant to changing the units of the parameters, i.e. to rescaling of the form 
\begin{equation}
    \partial_\mu S \mapsto w_\mu \partial_\mu S\label{eq:rescaling}.
\end{equation}

Each of these three optimality criteria depend nonlinearly on the state $\ket{E^{\mathrm{in}}}$. Therefore, gradient descent optimization is necessary to find the maximum information states. 

\subsubsection{Normalized Trace}
The trace of the Fisher matrix can also be used as an optimality criterion, which should not be confused with A-optimality. The trace has as benefit that the expression can be written as an inner product, which means that it can be maximized using an eigendecomposition. However, as mentioned in section \ref{sec:DAE-optimality} this criterion suffers from "winner takes all" tendency. To remedy this, we propose the following criterion. Define $f_\mu^{max}$ as the largest eigenvalue of $F_{\mu\mu}$. The normalized trace is then defined by
\begin{equation}
    \Tr _N\FI=\bra{E^{\mathrm{in}}}\pqty\bigg{ \sum_\mu \frac{ F_{\mu\mu} }{ f  ^{\mathrm{max}}_\mu } }\ket{E^{\mathrm{in}}}.
\end{equation}
By assigning weights to the entries in the trace, 
The maximum information state is then found by selecting the eigenvector with largest eigenvalue. Compared to optimization with gradient descent, this optimization is fast, deterministic and yields a complete set of orthogonal eigenvectors. The normalized trace can be seen as a rescaling \eqref{eq:rescaling} applied to the regular trace.

\begin{table}[tbp]
\begin{tabular}{lll}
Optimization goal & Expression                                                                  & Optimization Method \\ \hline
Scalar CRLB       & $\mathcal J_{\mu\mu}$                                                                    & Eigenvalue          \\
Joint CLRB        & $-\mathcal J_{\mu\mu}^{-1}$                                                               & Gradient Descent    \\
Normalized Trace  & $\bra{E}\pqty\bigg{ \sum_\mu \frac{ F_{\mu\mu} }{ f^{\mathrm{max}}_\mu } }\ket{E}$ & Eigenvalue \\
D-optimality      & $\mathrm{det}\mathcal J$                                                                 & Gradient Descent    \\
A-optimality      & $-\mathrm{tr}\mathcal J^{-1}$                                                             & Gradient Descent    \\
E-optimality      & $\mathrm{min}\{\lambda_1,\lambda_2,\dots\}$                                              & Gradient Descent    \\
                  &                                                                                          &                    
\end{tabular}
\caption{Different optimization goals to be maximized.}
\label{tab:optimization-goals}
\end{table}

\subsection{Nuisance Parameters}
A nuisance parameter (1) carries some non-zero variance, (2) correlates with the parameters of interest (3) is uninteresting to measure, i.e. there is no direct benefit to having a low CRLB on this nuisance parameter. It makes sense to sacrifice some precision on the nuisance parameter in order to achieve higher FI on the parameters of interest. But we have to be careful, because simply reducing the FI on the nuisance parameter can decrease performance on the parameter of interest if cross-information is present. We will look at optimality goals that take these effects into account.

\subsubsection{Partial Fisher Information}
Let us decompose the parameters into $\theta=(\poi,\nuis)^T$, where $\poi$ are the parameters of interest and $\nuis$ the nuisance parameters. The symbol $\nuis$ is fitting since it is the most annoying letter to write. The Fisher information matrix can then be written as a block matrix:

\begin{equation}
    \FI(\theta)=
        \begin{pmatrix} 
            \FI_{\poi\poi} & \FI_{\poi\nuis} \\
            \FI_{\nuis\poi} & \FI_{\nuis\nuis} 
        \end{pmatrix}.
\end{equation}
Since every principal submatrix of a positive semidefinite matrix is also positive semidefinite\cite{horn_matrix_2017}, we can immediately deduce from the joint Cram\'er-Rao bound \eqref{eq:matrix-CR} that
\begin{equation}
    \Sigma_{\poi\poi}\succeq \FI_\schur{\poi}{\nuis}^{-1}(\theta) ,\label{eq:marginal-CR}
\end{equation}
where $\FI_\schur{\poi}{\nuis}=\FI_{\poi\poi}-\FI_{\poi\nuis}\FI_{\nuis\nuis}^{-1}\FI_{\nuis\poi}$ is the Schur complement of the $\poi$ block, which is an explicit expression for the $\poi$-block of the inverse Fisher matrix. We will refer to $\FI_\schur{\poi}{\nuis}$ as the \textit{partial Fisher information}\cite{kotz_information_1992, bhapkar_conditioning_1989, van_trees_detection_2002, kay_fundamentals_2013}. Since the term $\FI_{\poi\nuis}\FI_{\nuis\nuis}^{-1}\FI_{\nuis\poi}$ is positive semi-definite, we can see that the presence of nuisance parameters can only decrease the Fisher information for the parameter of interest.

Here, we will investigate whether using the partial Fisher information combined with the previously mentioned optimality criteria will perform better for the parameters of interest. Judging from the form of the partial Fisher information, one would expect that using partial Fisher information as an optimization goal would do three things. It would increase Fisher information of the parameters of interest by the presence of the $\FI_{\poi\poi}$ term, decrease correlations via the $\FI_{\poi\nuis}$ term and also increase Fisher information on the nuisance parameter via the $\FI_{\nuis\nuis}^{-1}$ term, the latter two having a negative sign in front.

Another way to look at it is that for the parameters of interest, optimizing using the partial Fisher information matrix is identical to optimizing using the full Fisher information matrix. But in contrast, for the nuisance parameters direct optimization is removed. We have seen that multiparameter optimization is often a trade off. So by removing direct optimization on nuisance parameters, the procedure can direct Fisher information away from the nuisance parameter and towards the parameters of interest.

\subsubsection{Subblock Fisher information}
When dealing with nuisance parameters there is an even simpler approach compared to partial Fisher information: ignoring the nuisance parameter. This is equivalent to considering the subblock of the parameters of interest $\FI_{\poi\poi}$ as the Fisher matrix. Formally, it amounts to letting the Fisher information of the nuisance parameters go to infinity, or to consider the nuisance parameters as known. We will refer to this as \textit{subblock Fisher information}. The benefit of this approach is clear; measuring the scattering matrix takes time and if measuring it for the nuisance parameters does not give a clear improvement, you might as well leave them out. We will investigate in \ref{sec:results-nuisance} how subblock Fisher information compares to partial Fisher information.

\subsubsection{Naming Conventions}
In literature, partial Fisher information is also referred to as  horizontal information, profile information or efficient information\cite{fewster_information_2013}. It also goes unnamed sometimes. Subblock Fisher information is more formally the principal submatrix of the information matrix.

The inverse of the Fisher matrix can be viewed as the best possible covariance matrix, since it is the expected covariance matrix of efficient unbiased estimators when they exist. In the case of partial Fisher information, the corresponding covariance matrix is just the submatrix $\Sigma_{\poi\poi}$, which is also the marginal covariance of $\poi$. In other words, partial Fisher information is associated with the marginal covariance matrix. In a similar fashion, subblock Fisher information corresponds to $\Sigma_{\poi\poi}-\Sigma_{\poi\nuis}\Sigma_{\nuis\nuis}^{-1}\Sigma_{\nuis\poi}$, which is the partial covariance matrix . 

\begin{table}[tbp]
\begin{tabular}{ll}
Nuisance strategy           & Mathematical expression   \\ \hline
Full Fisher matrix          & $\FI_{\theta\theta}$      \\
Subblock Fisher information & $\FI_{\poi\poi}$          \\
Partial Fisher information  & $\FI_\schur{\poi}{\nuis}=\FI_{\poi\poi}-\FI_{\poi\nuis}\FI_{\nuis\nuis}^{-1}\FI_{\nuis\poi}$
\end{tabular}
\caption{Different nuisance parameter strategies. The parameters are decomposed as $\theta=(\poi,\nuis)^T$, where $\poi$ are the parameters of interest and $\nuis$ the nuisance parameters.\label{tab:nuisance-strategies}}
\end{table}

\section{Method}

\subsection{System Description\label{sec:system}}
To illustrate the multiparameter Fisher operator, we are simulating a 2D system of coupled dipoles\cite{bouchet_influence_2020, foldy_multiple_1945}. The system consists of a number of scatterers which have negligible size and are spaced far enough apart such that they can be approximated as point scatterers. We assume monochromatic illumination with wavelength $\lambda$ and scalar propagation such that polarization effects are ignored.

\begin{figure}
    \centering
        
        
    \includegraphics[width=0.49\linewidth]{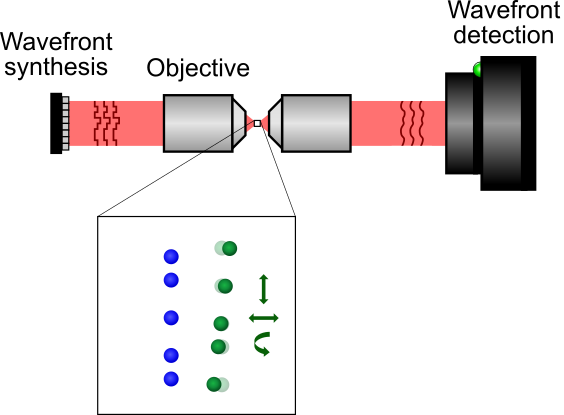}
    \caption{Schematic of the setup. Wavefront synthesis represents a device that can generate holograms with phase and amplitude control. Similarly, wavefront detection can be replaced with any device or setup that can detect holograms. In the inset the simulation box of size $8\lambda\times8\lambda$ is shown. The position and tilt of the rightmost row of scatterers (green) is to be estimated with respect to the left row (blue).}
    \label{fig:setup}
\end{figure}

We consider two columns of 5 scatterers each as shown in Figure \ref{fig:setup}. The positions of the rightmost row are parametrized by $(x, z, \phi)$, which represent horizontal translations, vertical translations and rotations respectively.  The positions of the scatterers were chosen as arbitrary sections of the Fibonacci chain in order to break any symmetries. This was done because we want to model two complex layers and not the particular symmetries of this system.

\begin{figure}
    \centering
        
        
    \includegraphics[width=0.985\linewidth]{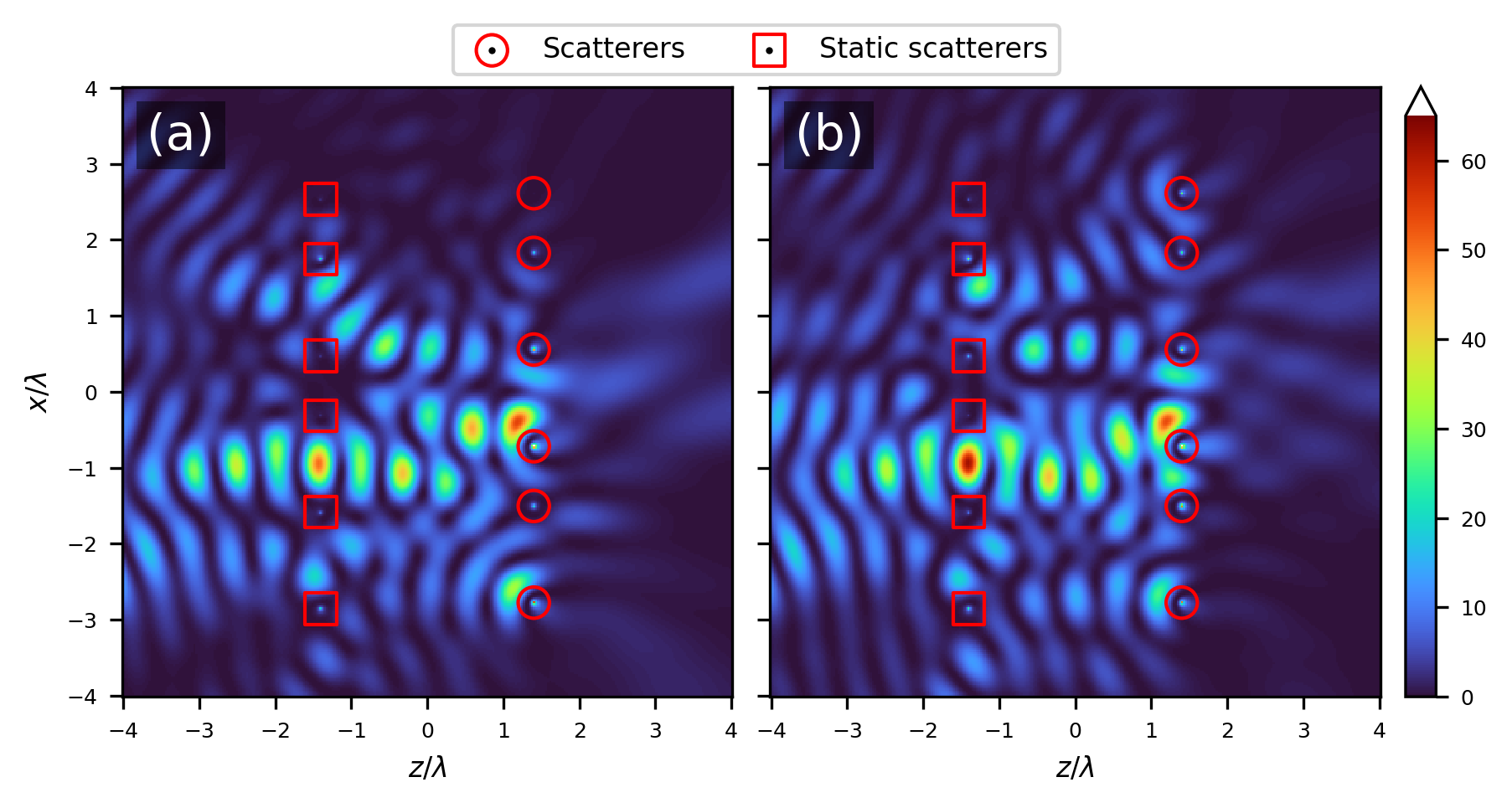}
    \caption{Intensity plots of maximum information states. (a) is optimized for the scalar Fisher information in $x$. (b) is optimized for the determinant of the Fisher matrix and allows for efficient estimation of all 3 parameters. Normalization is chosen such that a plane wave has intensity 1.}
    \label{fig:double_optimized}
\end{figure}

In Figure \ref{fig:double_optimized} the system is shown with two maximum information states. One is optimized for single parameter Fisher information and the other for multiparameter Fisher information.

\subsection{Scattering Matrix}
To calculate the scattering matrix, we must present the system with a number of flux normalized modes and then record the result in the same basis. We choose a basis of plane waves. Here it is important to ensure orthogonality of the basis modes. An overcomplete basis can result in overcounting of Fisher information if implemented naively. 

We choose a basis of plane waves with 45 modes. We choose 45 since in a 3D system we expect the square of the number of 2D modes, $45^2=2025$, which roughly corresponds to the number of modes in objective based setups\cite{pai_accurate_2020, bosch_measurements_2020, bouchet_maximum_2021}. In this paper we only look at the transmission matrix of the system, which is a subset of the full scattering matrix. We choose then the size of our simulation to match this number of modes and an NA of 0.95. This size is larger than the $8\lambda$ size used for plotting.

To calculate the Fisher tensor, the partial derivatives of the scattering matrix with respect to each parameter have to be calculated \eqref{eq:MPFI}. Here they are calculated using a discrete symmetric difference,
\begin{equation}
    \pdv{S}{\theta_\mu}\approx \frac{
            S(\theta+e_\mu\delta\theta)-S(\theta - e_\mu\delta\theta) 
        }{
            2\delta\theta
        },
\end{equation}
where $e_\mu$ is the unit vector in direction $\mu$. Here, symmetric difference was used for accuracy, but since calculating a scattering matrix can be expensive, a forward difference can be considered in order to reuse $S(\theta)$ multiple times.

\section{Results}
Here we will discuss the following questions using our framework. How do the CRLBs of multiparameter optimization goals compare to the scalar case? And what is the best strategy to deal with nuisance parameters? We discuss the former in Section \ref{sec:multiparam_vs_scalar} and the latter in Section \ref{sec:results-nuisance}.
    
\subsection{Multiparameter vs. Scalar\label{sec:multiparam_vs_scalar}}
\begin{figure}
    \centering
    \includegraphics[width=\linewidth]{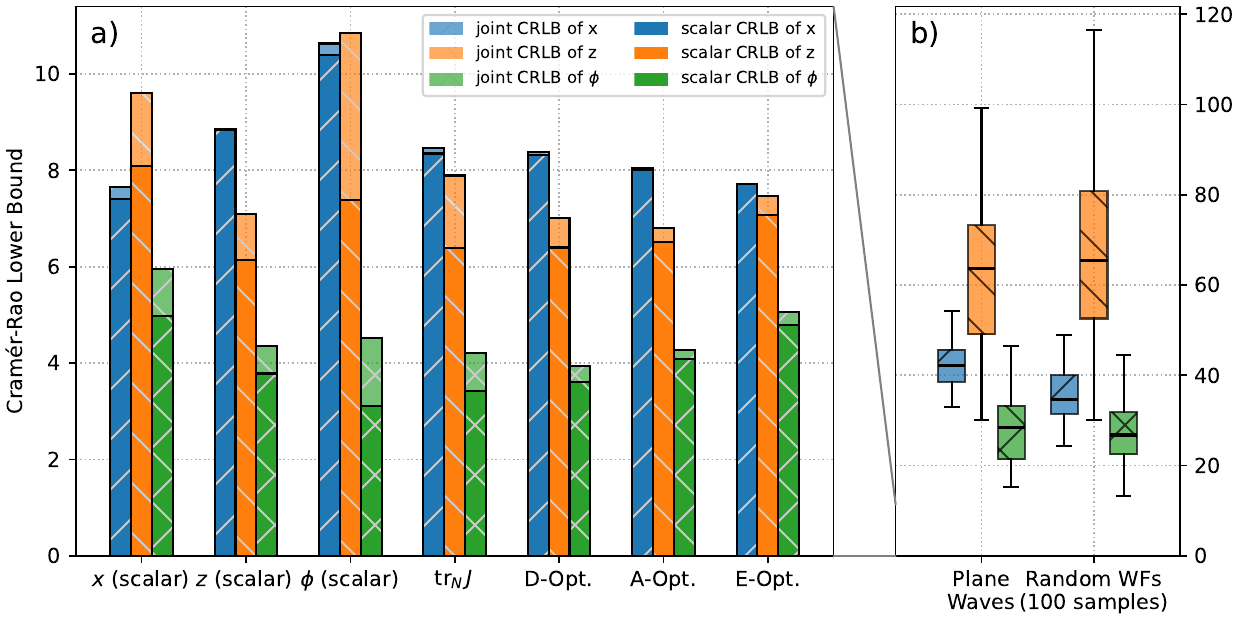}
    \caption{Cram\'er-Rao Lower Bounds (CRLB) of our 3 different parameters expressed as standard deviations. Lower is better. a) Shown for the optimality criteria that are discussed. See Table \ref{tab:optimization-goals}. b) For comparison, quartile boxplots are shown for a range of plane waves and an ensemble of 100 random wavefronts. The units of $x$ and $z$ are $10^{-3}$ wavelengths and that of $\phi$ is $10^{-3}$ radians. A photon budget of $500$ photons was used.}
    \label{fig:crlb_barplot.png}
\end{figure}

In Figure \ref{fig:double_optimized} we can see the results of optimizing for scalar Fisher information compared to optimizing a multiparameter optimality criterion, the determinant of the Fisher matrix. Both plots look similar at first sight, but one can notice that for the determinant more intensity is directed towards the top scatterers, which improves the Fisher information for rotation. We quantify this in Figure \ref{fig:crlb_barplot.png}, where both the scalar and joint CRLBs are shown for the optimality criteria that are discussed.

Firstly, note that the difference between scalar and joint CRLB is caused by off-diagonal terms in the Fisher matrix: for a diagonal matrix the scalar and joint CRLBs are equal. Unsurprisingly, we can see that the difference is the smallest for D-, A- and E-optimality. These design goals have knowledge of the off-diagonal terms and can optimize for them. As a result, the joint CRLBs tend to be lower.

We can also see that some multiparameter goals \textit{outperform} the scalar optimization goals. For example, the scalar optimization for $\phi$ is outperformed by all multiparameter goals except for E-optimality. This may seem surprising since scalar optimization is provably optimal, but this only holds when looking at the scalar CRLBs. The joint CRLB on a parameter can be improved by either reducing cross-information or by increasing the information on the other parameters, which makes it possible for the multiparameter goals to outperform the scalar optimization goals.

We should also note that even the worst maximum information states perform better than the best plane wave or random wavefront. In Bouchet et al. \cite{bouchet_maximum_2021} a 300-fold enhancement in Fisher information over the best plane wave is quoted, while here we see only a roughly $10\times$ enhancement. We suspect that this most this discrepancy can be explained by the larger number of modes inherent to 3D setups. In a more similar 2D geometry \cite{bouchet_influence_2020}, we see a similar enhancement.

    \subsection{Nuisance parameters\label{sec:results-nuisance}}
\begin{figure}
    \centering
    \includegraphics[width=\linewidth]{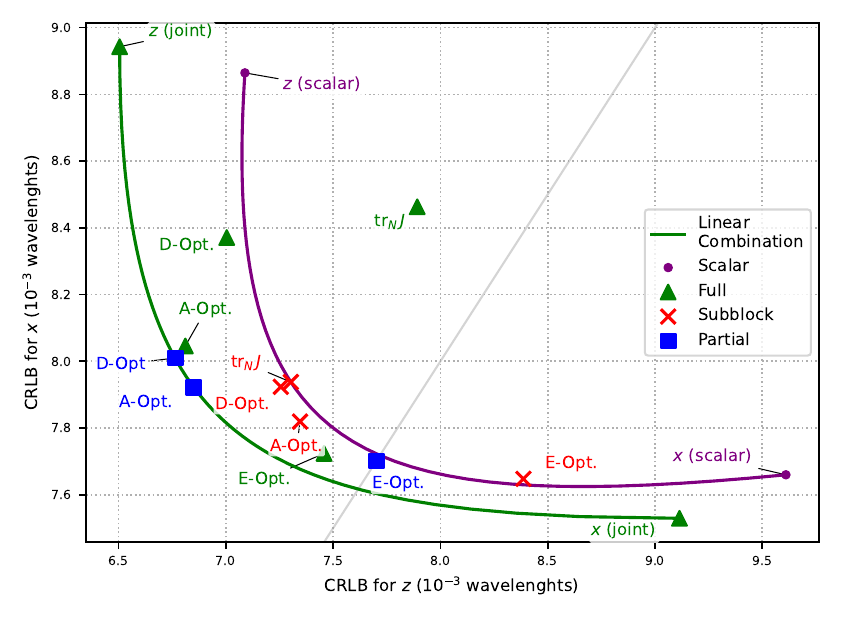}
\captionsetup{singlelinecheck=off}
    \caption[]{
    Cram\'er-Rao Lower Bounds (CRLB) for the two parameters of interest $z$ and $x$. Rotation is considered a nuisance parameter.  The labels indicate the optimization criteria; A-, D- or E-optimality, normalized trace and scalar CRLB. See also Table \ref{tab:optimization-goals}. The different symbol indicate strategy to deal with nuisance parameters. See also Table \ref{tab:nuisance-strategies}. A photon budget of $500$ photons was used.
    \label{fig:pareto_plot_nuisance.png}
    }
\end{figure}

To investigate the best way to deal with nuisance parameters, we will look at the joint CRLBs for $x$ and $z$ while considering $\phi$ as nuisance. In Figure \ref{fig:pareto_plot_nuisance.png} we can see the various optimality criteria discussed in \ref{sec:optimality-criteria} section are plotted. Also shown in green is an optimality criterion that consists of the linear combination $(1-\alpha)J^{-1}_{zz}+\alpha J^{-1}_{xx}$ for $\alpha\in[0,1]$. This curve forms a pareto-optimal frontier: no point can be better than this curve. Here a point is considered better than another point if \textit{all} of the CRLBs are better than that of the other point; the point Pareto dominates the other point. See the "weighting method" in Miettinen\cite{miettinen_nonlinear_1998}. Computing this curve requires considerable computational effort, which is why the usual optimality criteria are preferable in a setting with time constraints.

We should note that the appearance of Figure \ref{fig:pareto_plot_nuisance.png} depends strongly on the configuration of the dipoles, so we should be careful not to generalize too quickly. Having said that, there are some things we can conclude.

We can notice that for D- and A-optimality the points using partial Fisher information consistently land on the Pareto frontier. This gives confidence that partial Fisher information is an effective way to deal with nuisance parameters. For E-optimality this is not the case. We suspect that for E-optimality there is so much focus on having equal CRLBs for all parameters that it goes at the cost of achieving optimal results.

Also, we noticed that subblock Fisher information did either better or the same compared to using the full Fisher matrix, in the Pareto dominant sense. In Figure \ref{fig:pareto_plot_nuisance.png}, subblock Fisher information performs the same as using full Fisher information for all points except for normalized trace, where it does better. It would make sense that subblock Fisher information would perform better than full, since the optimization process does not have to waste its limited resources on the nuisance parameter. When the cross-information is large, which is the case in this geometry, it is possible to decrease the CRLB of a parameter of interest by simultaneously decreasing the cross-information and increasing the Fisher information on the nuisance parameter.

\section{Discussion and Conclusion}
We have introduced a multiparameter Fisher tensor that allows to find maximum information states by measuring the scattering matrix, which is a generalization of single parameter maximum information states. Using the Fisher tensor, the Fisher matrix can be easily calculated for any input state. The diagonals of the inverse Fisher matrix form the joint Cram\'er-Rao Lower Bounds, which represent the smallest variance possible while estimating a parameter limited by photon noise. One can optimize the joint CRLBs for the input state to get maximum information states in the multiparameter setting. One can also optimize for a linear interpolation of these joint CRLBs to get a line that is Pareto optimal. To avoid computing the entire line, D- or A-optimality evaluated on partial Fisher information are optimization goals that likely give you a solution on this optimal line. E-optimality will also give you a solution close to this optimal line, but seems to prioritize having equal CRLBs over achieving Pareto optimality. We have also introduced the \textit{normalized trace} optimality criterion, which can be optimized efficiently and provably optimal by eigenvalue decomposition. It is also more balanced compared to the trace of the Fisher matrix. However, the resulting states are not Pareto efficient. It is still an open question if it is possible to construct an optimality criterion which can not only be optimized by eigenvalue optimization, but is also Pareto efficient or sufficiently close.

We have also studied strategies for dealing with nuisance parameters. By using partial Fisher information it is possible to find Pareto efficient states in the parameters of interest, while not wasting too much resources on the nuisance parameter. In the framework of the Fisher information, it is still beneficial to have Fisher information on the nuisance parameter when cross-information is present. This is why subblock Fisher information performs worse compared to partial Fisher information. However, the penalty of leaving the nuisance parameter out of the optimization can be worth it depending on the cost of measuring the nuisance parameter's Fisher information.

One of the limitations of this work is that in order to experimentally measure the partial derivatives in the Fisher tensor, one needs to be able to apply small but known shifts to the parameters, which could be hard to realize. Also the ability to measure a scattering matrix is required. We should note that measuring the scattering matrix can take some time. However, the rank of the partial derivative of the scattering matrix is often considerably smaller than the scattering matrix itself, since only modes that interact with the parameter of interest survive. This opens up possibilities to save on measuring time by only measuring the non-zero part of this matrix.

As a final discussion point, we would like to discuss the possibility of time-sharing. Instead of using one illumination to estimate both parameters, it is also possible to first illuminate with a state optimized for one parameter and then illuminate with a state optimized for the other parameter. We refer to the latter as time-sharing. Let $E_z$ and $E_x$ be two states optimized for the joint CRLBs of $z$ and $x$ respectively. Since Fisher information is additive, the resulting Fisher matrix is simply $(1-t)\FI(E_z)+t\FI(E_x)$ where $t\in[0,1]$ is the fraction of the exposure time dedicated to state $E_x$. The resulting CRLBs are \textit{worse} than those obtained from the optimality criterion $(1-\alpha)J^{-1}_{zz}+\alpha J^{-1}_{xx}$. So at first glance time-sharing does not outperform single illumination. However, it is also possible to optimize the expression $(1-t)\FI(E_z)+t\FI(E_x)$ jointly for two different states $E_z, E_x$. Conceivably, this could improve the CRLBs by having the off-diagonals of the two matrices cancel. This question of efficiency using time-sharing is a potential future research direction for multiparameter maximum information states.

\begin{backmatter}
    \bmsection{Funding}
         This research was supported by the Topconsortia voor Kennis en Innovatie (TKI) programme under Grant PPS\_2023\_030 or 00PPS340; ASML
    \bmsection{Acknowledgment}
        
We would like to thank Omar al Gawhary, Jan de Graaff, Zili Zhou and Paul Urbach for helpful discussions on Fisher information. We would also like to thank Stefan van den Hoven and Xiaomeng Sui for discussions on time-sharing.

    \bmsection{Disclosures}
        B.V., J.S.: ASML (F). A.P.M.: ASML (RS).
    \bmsection{Data availability}
        The code and data used to generate the results in this paper are available on GitHub and have been permanently archived on Zenodo at \url{https://doi.org/10.5281/zenodo.22255379}

    \clearpage
    \startappendix{Supplemental document}
        \section{Multiparameter Quantum Fisher Information\label{ap:MQFI}}
A quantum system can be represented by a density matrix $\rho$,
\begin{equation}
    \rho=\sum_{i=1}^s p_i\ket{\psi_i}\bra{\psi_i},\label{eq:density_matrix}
\end{equation}
where $\sum_ip_i=1$ and $\{\ket{\psi_i}\}$ forms an orthonormal basis.
When $s=1$ the state is pure and any variance in measurements will be due to quantum effects. When $s\geq 2$, any variance will be a combination of quantum effects and experimental uncertainty. Using the density matrix we can calculate the Quantum Fisher Information Matrix (QFIM), which produces a Cram\'er-Rao bound that includes  both effects. The multiparameter QFIM is given by\cite{liu_quantum_2020}
\begin{equation}
    \QFI_{\mu\nu}=\tfrac 1 2\Tr(\rho\{L_\mu,L_\nu\}),\label{eq:MQFI}
\end{equation}
where $\{L_\mu,L_\nu\}=L_\mu L_\nu+L_\nu L_\mu$ is the anticommutator and where $L_\mu$ is the symmetric logarithmic derivative with respect to parameter $\theta_\mu$, defined implicitly by
\begin{equation}
    \partial_\mu \rho=\tfrac 1 2(\rho L_\mu+L_\mu\rho).
\end{equation}
Using the definition of the density matrix \eqref{eq:density_matrix}, we can expand the QFIM. By following the steps in\cite{liu_quantum_2014} closely, we obtain

\begin{equation}  
\begin{split}  
    \QFI_{\mu\nu}= \frac 1 2\bigg(&
    \sum_{i=1}^{s}\frac{\partial_\mu p_i\partial_\nu p_i}{p_i}
    + \sum_{i=1}^s 4p_i \pqty{ \braket{\partial_\mu \psi_i}{\partial_\nu \psi_i} - \braket{\partial_\mu \psi_i}{\psi_i} \braket{ \psi_i}{\partial_\nu\psi_i} }\\
    &- \sum_{i\neq j}^{s}\frac{ 8p_ip_j}{p_i + p_j}\braket{\partial_\mu \psi_i}{\psi_j} \braket{ \psi_j}{\partial_\nu\psi_i} 
      \bigg)+ (\mu\leftrightarrow\nu ).\label{eq:quantum_fisher_expanded}
\end{split}
\end{equation}

The first term in \eqref{eq:quantum_fisher_expanded} shows the classical contribution and only includes experimental uncertainty. The second term is the quantum contribution and the final term is a penalty due to mixing. The penalty is zero for pure states and is largest for near-uniform mixing.
\subsection{Coherent States}
We are now tasked with finding the QFIM for coherent states. Let us decompose the classical outgoing field into modes $E=\sum_kE^{out}_k$. For coherent illumination these modes will be populated by coherent or Glauber states $\ket{\alpha_k}$, labeled by the complex number $\alpha_k$. This complex number is then equal to the classical mode coefficient, i.e. $E_k=\alpha_k$. Let us look at the density matrix for a single such mode
\begin{equation}
    \rho=\ket{\alpha_k}\bra{\alpha_k}.
\end{equation}
We will drop the subscript $k$ for now. The QFIM \eqref{eq:quantum_fisher_expanded} for such a pure state then reduces to
\begin{equation}
    \QFI_{\mu\nu}=2\pqty{ \braket*{\partial_\mu \alpha}{\partial_\nu \alpha} - \braket*{\partial_\mu \alpha}{\alpha} \braket*{ \alpha}{\partial_\nu\alpha} } + (\mu\leftrightarrow\nu).
\end{equation}
To calculate the derivatives of these states, we opt to use the displacement operator
\begin{equation}
    \ket{\alpha}=D(\alpha)\ket{0}=\exp(X)\ket{0},\label{eq:glauber_definition}
\end{equation}
where we introduced $X\equiv\alpha a^\dagger-\alpha^* a$ for brevity and $a^\dagger, a$ are the creation and annihilation operators of the k-th mode. We can then use the following formula for the derivative of the matrix exponential\cite{hall_lie_2015}:
\begin{equation}
    \pdv{}{\theta_\mu}e^{X(\theta)}=e^{X(\theta)}\pqty{
        \partial_\mu X - 
        \frac{1}{2!}\comm{X}{\partial_\mu X} + 
        \frac{1}{3!}\comm{X}{\comm{X}{\partial_\mu X}} - \dots
        }.
\end{equation}
A quick calculation shows 
\begin{equation}
    [X,\partial_\theta X] =2i\Im(\alpha\partial_\theta\alpha^*).\label{eq:X_commutator}
\end{equation}
Since the commutator is just a number, any higher order commutators will vanish. Equation \eqref{eq:X_commutator} allows us to work out the following expressions
\begin{align}
\ket{\partial_\theta\alpha} &= e^{X(\theta)}\pqty{ \partial_\theta X - i \Im(\alpha\partial_\theta\alpha^*)}\ket 0\\
    \braket*{\partial_\mu\alpha_k}{\partial_\nu\alpha_k}&=\partial_\mu\alpha^*_k\partial_\nu\alpha_k + \Im(\alpha_k \partial_\mu\alpha_k^*)\Im(\alpha_k \partial_\nu\alpha_k^*)\\
    \braket*{\partial_\mu\alpha_k}{\alpha_k}&=-i\Im(\alpha_k \partial_\mu\alpha_k^*).
\end{align}
Finally, we can write the quantum Fisher information matrix for coherent states as
\begin{equation}
    \QFI_{\mu\nu}=2\sum_k (\partial_\mu\alpha^*_k\partial_\nu\alpha_k + \partial_\nu\alpha^*_k\partial_\mu\alpha_k).\label{eq:glauber_MQFI}
\end{equation}
Here we have used $\rho=\bigotimes_k \ket{\alpha_k}\bra{\alpha_k}$ and $\QFI(\bigotimes_k \rho_k)=\sum_k \QFI(\rho_k)$ to generalize to multiple modes\cite{liu_quantum_2020}.

Using an identical reasoning to the supplementary of \cite{bouchet_maximum_2021}, we can relate the QFIM in \eqref{eq:glauber_MQFI} to the scattering matrix. Recall that $\alpha_k=E_k^\mathrm{out}$, where $E_k^\mathrm{out}$ is just the $k$-th coefficient of the classical output field, i.e $E_k^\mathrm{out}=\braket{k}{E^\mathrm{out}}$. Here the bras and kets are defined in the Hilbert space of classical modes, not to be confused with equation \eqref{eq:glauber_definition} where the Hilbert space is that of the number states of the $k$-th mode.
For a linear optical system, the output states are related to the input states via the scattering matrix $S$, i.e. $\ket{E^{\mathrm{out}}}=S\ket{E^{\mathrm{in}}}$. This allows equation \eqref{eq:glauber_MQFI} to be written in terms of the scattering matrix and its input states
\begin{equation}
    \QFI_{\mu\nu}=2 \bra*{E^{\mathrm{in}}}\partial_\mu S^\dagger\partial_\nu S + \partial_\nu S^\dagger\partial_\mu S\ket*{E^{\mathrm{in}}}.\label{eq:QFI_scattering_matrix}
\end{equation}

        \section{Classical Measurement\label{ap:MFI}}
In equation \eqref{eq:QFI_scattering_matrix} we have shown an expression for the quantum Fisher information. It is now the question whether there exists a classical measurement scheme that achieves this bound. Similar to in \cite{bouchet_maximum_2021}, we propose a homodyne detection scheme where we record the intensity of the signal added to some reference beam with identical frequency, i.e.
\begin{equation}
    I_k(\theta)=\abs{E^{\mathrm{out}}_k(\theta)+E^{\mathrm{ref}}_k}^2.
\end{equation}
If we treat the light semiclassically, the detected intensity in mode $k$ will follow a Poisson distribution
\begin{equation}
    p(\vec n;\theta)=\prod_k\frac{(I_k(\theta))^{n_k} e^{-I_k(\theta)}}{n_k!},
\end{equation}
where $n_k$ is the detected number of photons in mode $k$ and the expected value of $n_k$ is $I_k$. The Fisher matrix associated with this distribution is given by
\begin{equation}
    \FI_{\mu\nu}(\theta)=\sum_k\frac{(\partial_\mu I_k )(\partial_\nu I_k)}{I_k}.
\end{equation}
We can write this as 
\begin{equation}
    \FI_{\mu\nu}(\theta)=\sum_k\frac{
    4\Re[(E^{\mathrm{out}}_k + E^{\mathrm{ref}}_k)(\partial_\mu E^{\mathrm{out}}_k)^*] 
    \Re[(E^{\mathrm{out}}_k + E^{\mathrm{ref}}_k)(\partial_\nu E^{\mathrm{out}}_k)^*]}
    {\abs{E^{\mathrm{out}}_k + E^{\mathrm{ref}}_k}^2}
\end{equation}
Conveniently, this expression can be seen as a product of two vector projections since
\begin{equation}
\frac{\Re[(E^{\mathrm{out}}_k + E^{\mathrm{ref}}_k)(\partial_\mu E^{\mathrm{out}}_k)^*] }
{\abs{E^{\mathrm{out}}_k + E^{\mathrm{ref}}_k}} =
\frac{(\vec E^{\mathrm{out}}_k + \vec E^{\mathrm{ref}}_k)\cdot (\partial_\mu \vec E^{\mathrm{out}}_k) }
{\abs{\vec E^{\mathrm{out}}_k + \vec E^{\mathrm{ref}}_k}},
\end{equation}
where $\vec E$ is the 2D vector constructed from the complex number $E$. We can then simplify further by picking a high intensity reference beam $E^{\mathrm{ref}}_k\gg E^{\mathrm{out}}_k$, which yields the straightforward expression
\begin{align}
    \frac{\vec E^{\mathrm{ref}}_k\cdot \partial_\mu \vec E^{\mathrm{out}}_k }
{\abs{\vec E^{\mathrm{ref}}_k}}=\abs{\partial_\mu  E^{\mathrm{out}}_k}\cos(\arg (E^{\mathrm{ref}}_k)-\arg (\partial_\mu E^{\mathrm{out}}_k)).
\end{align}
For a single parameter, $\FI_{\mu\mu}$ is easily maximized by picking the argument of the reference beam to be equal to the argument of $\partial_\mu E^{\mathrm{out}}_k$ for each mode. In that case the Fisher information for this measurement scheme recovers the well-known $\FI_{\mu\mu}=\sum_k 4\abs{\partial_\mu E^{\mathrm{out}}_k}^2$. However, in the multiparameter case this is not so straightforward, since optimizing the reference angle for one parameter might result in zero information for another parameter. To still get a useful bound, we can use a measurement protocol discussed in Bouchet\cite{bouchet_fundamental_2021}. Instead of illuminating using one fixed reference angle, we can cut our measuring time into $N\geq 3$ intervals and illuminate each interval with a reference angle equal to $2\pi j/ N,$ with $j=1\dots N$. This will result in Fisher information that is exactly half of the quantum Fisher information, both for the diagonals and off-diagonals of the matrix since
\begin{align}
    \sum_{j=1}^N\cos(2\pi j/N-\phi_\mu)\cos(2\pi j/N-\phi_\nu)=
    \frac 1 2\sum_{j=1}^N \pqty{
    \cancel{\cos(4\pi j/N)}+\cos(\phi_\mu-\phi_\nu)}
\end{align}
where $\phi_\mu=\arg (\partial_\mu E^{\mathrm{out}}_k)$.
So using this protocol the Fisher information matrix is given by
\begin{align}
    \FI_{\mu\nu}(\theta)&=\sum_k (\partial_\mu  E^{\mathrm{out}}_k )^*\partial_\nu  E^{\mathrm{out}}_k + (\partial_\nu  E^{\mathrm{out}}_k )^*\partial_\mu  E^{\mathrm{out}}_k\\
    &=\bra{E^{\mathrm{in}}}\partial_\mu S^\dagger\partial_\nu S+ \partial_\nu S^\dagger\partial_\mu S\ket{E^{\mathrm{in}}},\label{eq:hetero_FI}
\end{align}
which is exactly a factor of one-half times the quantum Fisher matrix \eqref{eq:QFI_scattering_matrix}. We should note that this factor half penalty can be attributed to the fact that we have switched to doing heterodyne detection, since both quadratures are estimated at the same time. 

One possible way of restoring homodyne efficiency is by using a perfect programmable mode splitter. For each parameter $\theta_\mu$ present, a splitter arm will be added to select the mode $\partial_\mu E^{\mathrm{out}}$. In each splitter arm, a homodyne detector is used to extract maximum Fisher information. This protocol will then extract a Fisher matrix with homodyne efficiency, i.e. $\FI_{\mu\nu}(\theta)=2\bra{E^{\mathrm{in}}}\partial_\mu S^\dagger\partial_\nu S+ \partial_\nu S^\dagger\partial_\mu S\ket{E^{\mathrm{in}}}$. Any mode overlap between the modes for two parameters $\mu_1$ and $\mu_2$ is fine, since the Fisher information associated with overlap will end up in the mode associated with either $\mu_1$ or $\mu_2$. However, a perfect programmable mode splitter for many modes has never been demonstrated and so for multiparameter Fisher information we are practically limited to heterodyne efficiency.
        \section{Bias}
We are now interested in how the bias of maximum information states grows as a function of the nuisance parameter. Bias is simply the difference between the predicted and true value, $b=\hat \theta-\theta$, and it is a measure of accuracy. We will investigate this by assuming the field deviates linearly around some reference value as a function of the parameters\cite{bouchet_maximum_2021}. How far can we deviate from the reference value before the estimation breaks down? Before we answer this question we will first discuss how to estimate the parameters using knowledge of the partial derivatives.

\subsection{Multilinear Estimation}
We assume the output field takes on the form
\begin{equation}
    E(\theta)=E_0+H\theta+w.
\end{equation}
Here $\theta$ is a real vector of parameters. Without loss of generality, we assume that $\theta=0$ at the reference value. Also, $E(\theta)$ is a complex vector representing the output field and $E_0=E(0)$ is the field at the reference value. The complex matrix $H$ is the Jacobian matrix $H=\left.\frac{\partial E}{\partial \theta}\right|_{\theta=0}$ evaluated at the reference parameter value. To model quantum optics, the vector $w$ is taken to be a randomly distributed complex gaussian vector with zero mean and variance $\tfrac 1 2$ for the real and imaginary parts. Also, the field is normalized according to $|E^{\mathrm{in}}|^2=\bar n$, where $\bar n$ is the expected number of photons. This variance is consistent with jointly estimating both quadratures of a coherent state, where the variance of a single quadrature would be $\tfrac 1 4 $.

For estimation we can use the Best Linear Unbiased Estimator (BLUE), which has zero bias and a variance that saturates the CRLB\cite{kay_fundamentals_2013}.
The BLUE is  given by\cite{lang_best_2016}
\begin{equation}
    \hat\theta = \Re[H^\dagger H]^{-1}\Re[H^\dagger(E-E_0)].\label{eq:blue_complex}
\end{equation}
The BLUE has inverse covariance matrix $\Sigma^{-1}=\frac{2}{\sigma_{re}^2+\sigma_{im}^2}\Re[H^\dagger H]=2\Re[H^\dagger H]$, which is identical to the bound given by \eqref{eq:hetero_FI}. This expression scales linearly with the photon number since $H^\dagger H$ scales linearly with the norm of the input state.

\subsection{Bias as Function of Nuisance}
In Figure \ref{fig:bias_nuisance} we see how the bias varies as a function of nuisance. The nuisance parameter, rotation, is varied over $\pm0.02$ radians, which is about $\pm1.1$ degrees and corresponds to a displacement of 0.05 wavelengths on the outermost dipoles. We also tested the different strategies, full, subblock and scalar, but they showed very similar results. The bias functions are well approximated by $b_x(\phi)=6.3\phi^2-80\phi^3$ for $x$ and by $b_z(\phi)=3.7\phi^2-29\phi^3$ for $z$, where $\phi$ is in radians and the bias in wavelengths. To get a sense of scale, for light with wavelength 632.8 nm, a 1nm bias corresponds to $1.6\cdot10^{-3}$ wavelengths and the bias of $z$ stays just below this line over the window in Figure \ref{fig:bias_nuisance}.

\begin{figure}[htbp]
    \centering
    \includegraphics[width=0.5\linewidth]{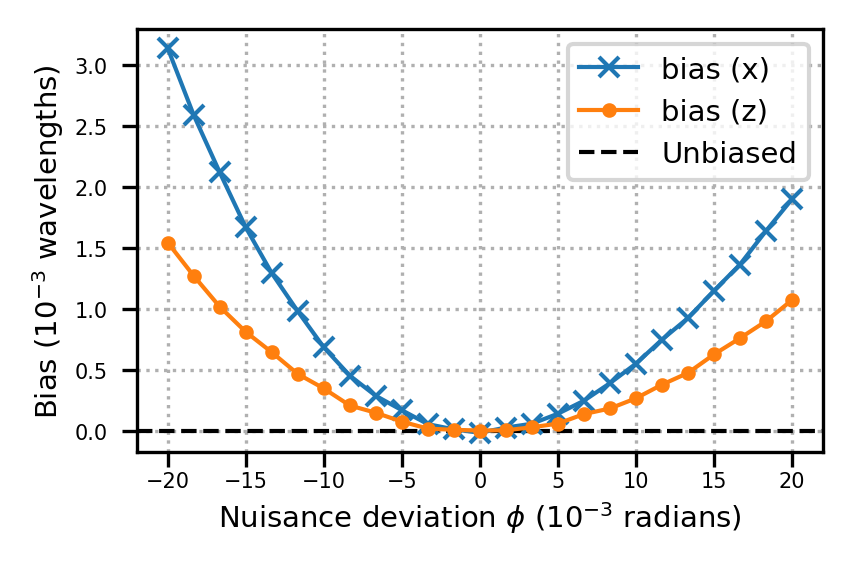}
    \caption{Bias as a function of nuisance parameter deviation. Bias is in units of wavelengths and can be negative. The used states are optimized for the partial Fisher information of D-optimality.}
    \label{fig:bias_nuisance}
\end{figure}



        
        
        

\end{backmatter}

\clearpage

\bibliography{references}

\end{document}